\documentclass[11pt,a4paper]{article}
\usepackage{epsfig}
\usepackage{amsmath}
\usepackage{amssymb}
\usepackage{verbatim}
\usepackage{bm}
\usepackage{color}
\usepackage{cancel}
\usepackage{ulem}
\usepackage{cite}
\usepackage{bbm}
\newcommand{\hwl}{\rule[0mm]{0mm}{4mm}}
\begin{document}
\thispagestyle{empty}
\begin{center}

{\LARGE\bf Natural Inflation and Low Energy Supersymmetry}\\[12mm]

{\large Rolf Kappl, Hans Peter Nilles, Martin Wolfgang Winkler} 
\\[6mm]
{\it Bethe Center for Theoretical Physics\\
and\\
Physikalisches Institut der Universit\"at Bonn\\
Nussallee 12, 53115 Bonn, Germany
}
\vspace*{12mm}
\begin{abstract}
Natural (axionic) inflation provides a well-motivated and
predictive scheme for the description of the early universe.
It leads to sizeable primordial tensor modes and thus a high
mass scale of the inflationary potential. Na\"{\i}vely this seems
to be at odds with low (TeV) scale supersymmetry,
especially when embedded in superstring theory. We show that
low scale supersymmetry is compatible with natural (high scale)
inflation. The mechanism requires the presence of two axions
that are provided through the moduli of string theory.

\end{abstract}
\end{center}
\clearpage

\section{Introduction}

The energy scale of inflation is directly related to the size
of tensor modes $r$ in the cosmological microwave background~\cite{Lyth:1996im}.
A large value of $r$ (as possibly indicated by recent observations~\cite{Ade:2014xna,Ade:2015tva}) 
would lead to a scale of the inflationary potential as 
large as the grand unification (GUT) scale 
$M_{\rm GUT} \sim 10^{16}\:\text{GeV}$. While the experimental situation
is not settled yet, such ``high scale inflation''  remains a 
realistic option. This scheme is of interest as the GUT-scale is 
not far from the Planck-scale and we might hope to obtain viable
insight in properties of gravitational interactions. High scale
inflation is challenging as our perturbative expansions based on 
the weak coupling limit might no longer be valid. We thus would have
to rely on a satisfactory ultraviolet (UV) completion (here assumed
to be superstring theory) and the symmetries of the system.

One of the prime candidates for high scale inflation is so-called
natural inflation~\cite{Freese:1990rb} based on an axionic
shift symmetry of the inflaton field. However, as any scheme of
inflation with sizeable tensor modes, natural inflation leads necessarily to
trans-Planckian excursions of the inflaton field (to obtain a
sufficient number of e-folds of inflation). This problem can be 
solved within the framework of monodromical motion of (several) 
axion fields~\cite{Kim:2004rp,Silverstein:2008sg} within superstring theory as a 
UV-completion.

In superstring theory there appears a new mass scale, the
breakdown scale of supersymmetry, represented by the gravitino
mass $m_{3/2}$. The question arises how this scale is related
to the energy scale of the inflationary potential. In the
presence of light scalar fields (as e.g. the string moduli) 
the process of inflation might be disturbed.

Na\"{\i}vely one might then 
postulate a supersymmetry breakdown scale larger than the scale of 
inflation. 
Unfortunately this would remove the
possibility of TeV-scale supersymmetry as a stabilizer for the
low mass of the Higgs boson. It would also imply
that supersymmetry does not protect
the flatness of the inflaton potential against radiative corrections.

In this paper we would like to discuss possible alternatives to
this na\"{\i}ve reasoning and see whether low scale (or unbroken)
supersymmetry is compatible with high scale inflation. Before 
moving to string theory, we shall
start with a discussion of the attempts to embed the scenario 
of natural inflation within the framework of the supergravity 
effective action (section~\ref{sec:natural}). This includes models of high scale supersymmetry breakdown~\cite{Kallosh:2007cc,Czerny:2014xja,Long:2014dta,Ben-Dayan:2014zsa} as well as supersymmetric models with stabilizer fields~\cite{Kallosh:2014vja}. 
In section~\ref{sec:string} we confront these
models
with the requirement of string theory embedding and 
the mechanism of moduli stabilization. We shall see that
a successful implementation of low scale supersymmetry is 
difficult
with just one axion field. 
Low scale supersymmetry
as well as the solution of the trans-Planckian excursion requires
a system with several fields, which are well motivated from 
string theory with its various moduli. Our main result is
given in section~\ref{sec:model} where we implement the axion alignment mechanism~\cite{Kim:2004rp}
within a supersymmetric model.\footnote{Implementations of aligned natural inflation in string theory have also been discussed in~\cite{Long:2014dta,Ben-Dayan:2014zsa,Gao:2014uha,Li:2014lpa,Ben-Dayan:2014lca,Abe:2014pwa,
Ali:2014mra,Abe:2014xja,Bachlechner:2014gfa,Li:2014unh,Garcia-Etxebarria:2014wla}.} This explicitly 
shows the compatibility of low scale supersymmetry within the
framework of high scale inflation.

\section{Natural Inflation in Supergravity}
\label{sec:natural}

The axion $\varphi$ is a well-motivated inflaton candidate. At the perturbative level, the Lagrangian is invariant under the transformation
\begin{equation}\label{eq:shift}
\varphi \rightarrow \varphi  + c\,.
\end{equation}
Hence, the axion corresponds to a flat direction in field space. At the non-perturbative level, the shift symmetry is (typically) broken to a discrete remnant and a periodic potential for the axion arises
\begin{equation}\label{eq:naturalinflation}
V = \Lambda^4 \left[1 - \cos\left(\frac{\varphi}{f}\right)\right]\,,
\end{equation}
where $f$ denotes the axion decay constant and $\Lambda$ is a mass scale related to the strength of the instanton effect which breaks the shift symmetry. This potential can accommodate inflation and, for $f\gg1$, gives rise to CMB observables $r$ and $n_s$~\cite{Freese:1990rb} in agreement with observation~\cite{Ade:2015lrj}. 

In supergravity, the axion $\varphi$ is part of a complex scalar field $\rho=\chi + i \varphi$, where $\chi$ denotes the saxion. In order to realize natural inflation, the K\"ahler potential $K$ is required to be invariant under the shift of the axion~\eqref{eq:shift}. The breaking of the shift symmetry is induced by a non-perturbative term $e^{a\rho}$ in the superpotential $W$ which leaves the Lagrangian invariant under certain discrete shifts.
In the following, we wish to distinguish natural inflation models with high scale supersymmetry breaking and models with a stabilizer field which preserve supersymmetry.

\subsection{Models with High Scale Supersymmetry Breaking}\label{sec:highm32}

The supergravity embedding of natural inflation requires that the scalar potential takes the simple form~\eqref{eq:naturalinflation} for the whole trans-Planckian excursion of the inflaton field. The shift symmetry in $K$ protects the inflationary potential against large supergravity corrections from the K\"ahler potential. Still, simple attempts to realize natural inflation with a single chiral superfield $\rho$ are doomed to fail. 
The global supersymmetry potential $|W_\rho|^2$ or the supergravity term $-3|W|^2$ alone can induce a cosine potential for the axion. However, the interference between both terms typically destroys this simple picture.

One way to solve this problem is to introduce a large constant $W_0$ in the superpotential which works as an order parameter~\cite{Ben-Dayan:2014lca}. In this case the derivative $W_\rho$ is suppressed compared to $W$. To be specific, let us consider a model defined by
\begin{subequations}\label{eq:nostabilizer}
\begin{align}
W &= W_0 + Ae^{-a\rho}\,,\label{eq:supp}\\
K&= \frac{(\bar{\rho}+\rho)^2}{4}\,.
\end{align}
\end{subequations}
Assuming $W_0\gg Ae^{-a\rho}$, we can expand the resulting potential in powers of $W_0$
\begin{equation}\label{eq:vhighm32}
V = W_0^2\; (2\chi^2 -3) e^{\chi^2} - W_0\; (3+ 2 a \chi - 2\chi^2)\,2 A \,e^{\chi^2-a\chi}\,\cos\left(a\varphi\right) + \mathcal{O}(W_0^0)\,.
\end{equation}
The saxion is stabilized at $\chi\simeq 1/\sqrt{2}$ with a large mass. After integrating out $\chi$,~\eqref{eq:vhighm32} agrees with the potential of natural inflation up to a constant shift in the vacuum energy $\mathcal{O}(-W_0^2)$.
Hence, an additional uplifting mechanism to adjust the vacuum energy is required. The simplest uplifting sector consists of one chiral superfield $\psi$ and a constant \(\mu\) with a linear superpotential term
\begin{equation}\label{eq:uplifting}
W_\text{up} = \mu^2 \psi\,,\\
\end{equation}
which has to be added to~\eqref{eq:supp}. The chiral superfield breaks supersymmetry and its F-term contribution to the potential can be tuned against $-3|W|^2$ in order to obtain a suitable Minkowski minimum.

Independent of the details of the uplifting mechanism, cancellation of the vacuum energy after inflation requires a gravitino mass $m_{3/2}\sim W_0 \gg H$, where $H$ denotes the Hubble scale during inflation. Since $H\sim 10^{-4}$ in natural inflation, supersymmetry must be broken at a very high scale. This is in analogy to the case of chaotic inflation, which can also be realized with a single chiral superfield at the price of introducing a high supersymmetry breaking scale~\cite{Buchmuller:2015oma}.

\subsection{Models with a Supersymmetric Ground State}
\label{sec:lowscale}

Let us now discuss models with a stabilizer~\cite{Kawasaki:2000yn,Kallosh:2010ug,Kallosh:2010xz,Kallosh:2014vja}. We start from the superpotential and K\"ahler potential
\begin{subequations}\label{eq:stabilizer}
\begin{align}
W &= m^2 X\, (e^{-a\rho} - \lambda)\,,\\
K&= \frac{(\bar{\rho}+\rho)^2}{4} + k(|X|^2)% -  \frac{|X|^4}{\Lambda^2}
\,,
\end{align}
\end{subequations}
where $X$ is the stabilizer field and $\rho$ contains the axion which is again protected by the shift symmetry in the K\"ahler potential. The K\"ahler potential of the stabilizer $k(|X|^2)$ is assumed to decouple $X$ from inflation via a quartic term~\cite{Kallosh:2010xz}.

The system has a supersymmetric ground state at $X=0$, $\rho=-\log(\lambda)/a\equiv \rho_0$. Setting $X$ to its minimum and $\rho=\rho_0 +\chi + i \varphi$, the scalar potential reads
\begin{equation}\label{eq:coscosh}
V=2 m^4 \,e^{(\rho_0 + \chi)^2-a (2\rho_0+ \chi)}\: \left[\cosh\left(a\chi\right) - \cos\left(a\varphi\right)\right]\,,
\end{equation}
As the minimum preserves supersymmetry, the masses of axion and saxion are identical, $m_\varphi=m_\chi$. 
However, slightly away from the minimum, the scalar potential in the direction $\chi$ becomes very steep due to the exponential factor $e^K$. During inflation, $\chi$ remains very close to the origin and~\eqref{eq:coscosh} effectively reduces to the well-known potential of natural inflation.

In this model, supersymmetry is restored at the end of inflation. Therefore, an additional sector of particles has to be added to account for supersymmetry breaking. A minimal example is comprised of a single chiral superfield with the superpotential
\begin{equation}
W_{\cancel{\text{susy}}}= \mu^2 \psi + W_0\,,
\end{equation}
as in the Polonyi model~\cite{Polonyi:1977}. Here, the constants $\mu$ and $W_0$ are unrelated to the scale of inflation. In case $\psi$ has a canonical K\"ahler potential, $\psi$ would get displaced during inflation leading to the well-known Polonyi problem~\cite{Coughlan:1983ci}. However, if the K\"ahler potential of $\psi$ contains an additional quartic term stemming from interactions with heavy fields, the minima of $\psi$ during and after inflation (almost) coincide. In this case the Polonyi problem is absent~\cite{Dine:1983ys,Coughlan:1984yk}. 

Hence, in natural inflation models with stabilizer, one can account for supersymmetry breaking by adding a simple hidden sector without causing cosmological problems. In particular, natural inflation models with stabilizer are consistent with $W_0\ll H$ and hence a low supersymmetry breaking scale $m_{3/2}\sim  \text{TeV}$. This appears favorable from a naturalness point of view and opens the possibility to discover supersymmetry at the Large Hadron Collider.

\section{Natural Inflation in String Theory}
\label{sec:string}

From the bottom-up perspective, the models discussed in the previous section yield a satisfactory supergravity embedding of natural inflation. However, these schemes crucially rely on the presence of the shift symmetry in the Lagrangian which is preserved at the perturbative level.
The question whether the shift symmetry survives quantum gravity effects requires knowledge of the UV theory, for which we consider string theory as the leading candidate.

String theory gives rise to a variety of axions descending from higher-dimensional p-form gauge fields. The string axions enjoy continuous shift symmetries which hold to all orders in perturbation theory~\cite{Wen:1985jz,Dine:1986vd}. Instanton effects, which will be discussed in section~\ref{sec:instantons}, break the shift symmetries down to discrete remnants.\footnote{Alternatively, flux-induced superpotentials could break the shift symmetry. This possibility has been investigated in the context of inflation in~\cite{Marchesano:2014mla,Hebecker:2014eua,Blumenhagen:2014nba}.} Hence, string theory contains the ingredients of a successful realization of natural inflation in the UV theory. Unfortunately, one cannot simply identify the axion introduced in the supergravity examples of section~\ref{sec:natural} with one of the string axions.

\subsection{Challenges}

An immediate problem is that in the supergravity examples, an axion decay constant $f\gg 1$ is required in order to match the constraints on CMB observables. In string theory, this poses a severe problem. The arguments of~\cite{Banks:2003sx,Svrcek:2006yi} indicate that trans-Planckian decay constants of string axions do never arise in a controllable regime of the theory (see however~\cite{Blumenhagen:2014gta,Grimm:2014vva}).

Leaving this aside, the discussed supergravity models do not address the problem of modulus (saxion) stabilization. Taking simple quadratic K\"ahler potentials as in section~\ref{sec:natural}, the saxion partner of the axion is stabilized within a very steep (exponential) potential. String-derived K\"ahler potentials of moduli fields are, however, of the form~\cite{Witten:1985xb}
\begin{equation}
 K=-\log(\bar{\rho} + \rho)\,,
\end{equation}
Given that $\rho$ only appears non-perturbatively in the superpotential, the logarithm implies that the scalar potential vanishes in the limit $\rho\rightarrow \infty$. The saxion is thus only protected by a potential barrier $\Delta V$ of finite height against destabilization towards run-away. With $\rho$ coupling to the vacuum energy through the K\"ahler potential, the saxion is generically displaced during inflation. If the inflationary vacuum energy exceeds $\Delta V$, the saxion is lifted over its potential barrier and never returns to the desired minimum of the scalar potential. This problem has been raised in the context of dilaton stabilization in the heterotic string~\cite{Buchmuller:2004xr} as well as K\"ahler moduli stabilization in type IIB~\cite{Kallosh:2004yh}. 

\begin{figure}[t]
\centering
\includegraphics[height=5.1cm]{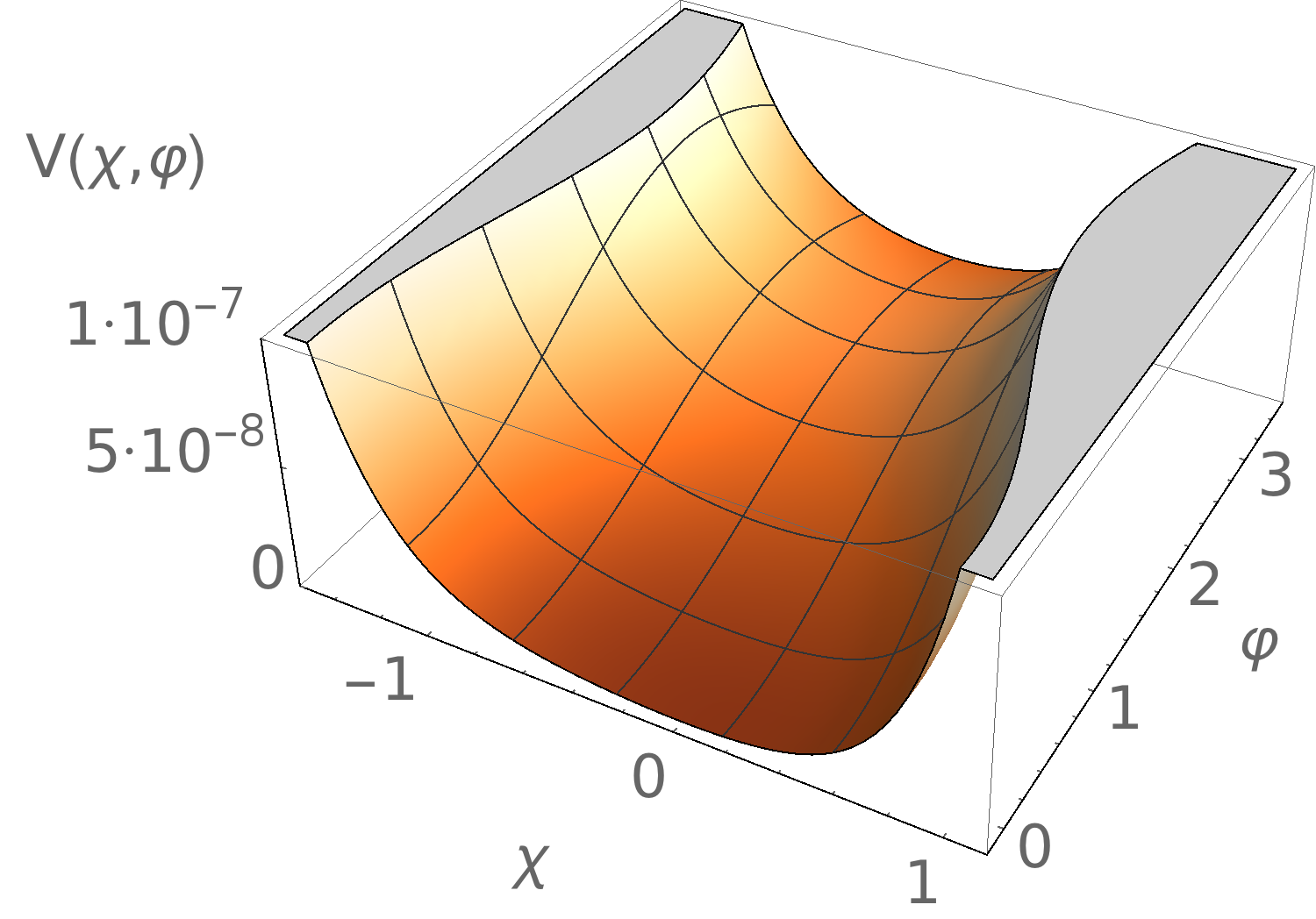}\hspace{3mm}
\includegraphics[height=5.1cm]{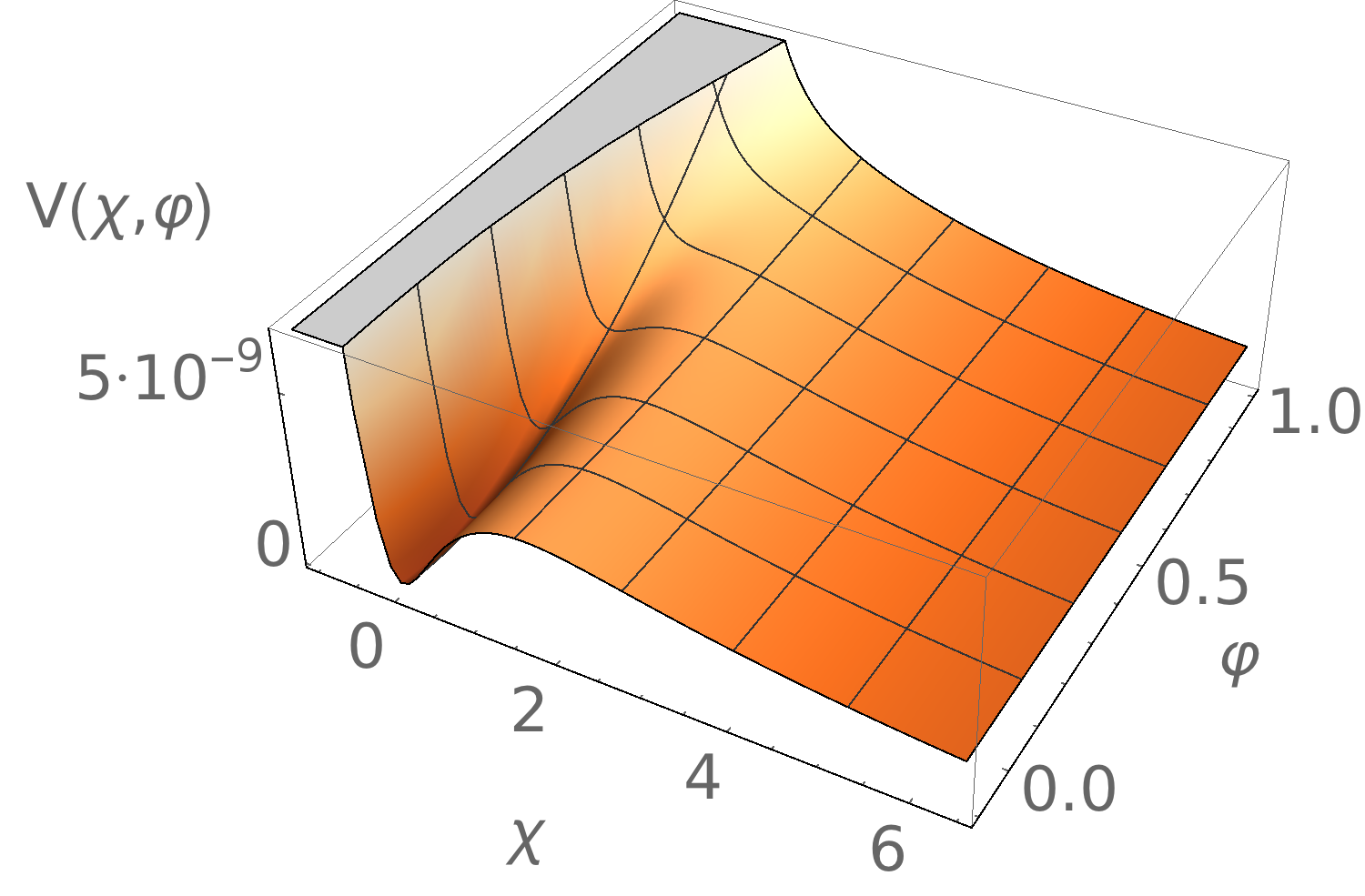}
\caption{Comparison of the scalar potentials~\eqref{eq:coscosh} and~\eqref{eq:modpot} arising from a quadratic K\"ahler potential (left panel) and from a logarithmic K\"ahler potential (right panel). In the logarithmic case, the saxion \(\chi\) is only protected by a small barrier from runaway.}
\label{fig:log3d}
\end{figure}

Assuming a logarithmic rather than a quadratic K\"ahler potential, the scalar potential changes from~\eqref{eq:coscosh} into
\begin{equation}\label{eq:modpot}
V=\frac{m^4 \,e^{-a(2\rho_0+\chi)}}{\rho_0+\chi} \left[\cosh\left(a\chi\right) - \cos\left(a\varphi\right)\right]
\,.
\end{equation}
In figure~\ref{fig:log3d} we compare the potentials~\eqref{eq:coscosh} and~\eqref{eq:modpot}. For the logarithmic K\"ahler potential, there is only a tiny barrier which protects the saxion from destabilization towards $\chi \rightarrow \infty$. One can easily verify that natural inflation cannot be realized with the logarithmic K\"ahler potential. The inflationary vacuum energy would lift $\chi$ over its potential barrier and the system would never reach the desired minimum of the potential. It is thus clear that a successful inflation model has to address the stabilization of moduli. 

In the next sections we wish to show that both problems: the trans-Planckian axion decay constant as well as modulus stabilization can be addressed using the axion alignment mechanism proposed by Kim, Nilles and Peloso (KNP)~\cite{Kim:2004rp}.

\subsection{Aligned Natural Inflation}

An effective axion decay constant \(f_{\text{eff}}\gg 1\) can be generated if the inflaton is a linear combination of two or more axions~\cite{Kim:2004rp,Tye:2014tja,Kappl:2014lra,Choi:2014rja}. We start with  a potential for two canonically normalized axions $\varphi_{1,2}$
\begin{equation}
V = \Lambda_1^4 \left[1 - \cos\left(\frac{\varphi_1}{f_1}+\frac{\varphi_2}{g_1}\right)\right]+\Lambda_2^4\left[1-
\cos\left(\frac{\varphi_1}{f_2}+\frac{\varphi_2}{g_2}\right)\right]\,,
\end{equation}
with all fundamental decay constants \(f_i,g_i<1\). For simplicity, we assume \(\Lambda_2^4\gg \Lambda_1^4\) such that the linear combination  \(\tilde{\varphi}\propto\frac{\varphi_1}{f_2}+\frac{\varphi_2}{g_2}\) is heavy, while the orthogonal combination \(\varphi\propto\frac{\varphi_1}{g_2}-\frac{\varphi_2}{f_2}\) remains light. For perfect alignment (\(\delta=0\)) of the axion decay constants, $\varphi$ corresponds to a flat direction, where
\begin{equation}
\delta=\frac{1}{f_1g_2}-\frac{1}{g_1f_2}\,.
\end{equation}
If we introduce a small misalignment \(\delta\neq 0\) the flat direction is lifted and we find a potential with an effective decay constant \(f_{\text{eff}}\)
\begin{equation}
V \simeq \Lambda_1^4 \left[1 - \cos\left(\frac{\varphi}{f_{\text{eff}}}\right)\right]\,,\qquad
f_{\text{eff}}=\frac{\sqrt{\frac{1}{f_2^2}+\frac{1}{g_2^2}}}{\delta}\,.
\end{equation}
For small \(\delta\) this effective decay constant can naturally be trans-Planckian as required in a successful model of natural inflation.

\subsection{Origins of Axion Potentials in String Theory}\label{sec:instantons}

Previous attempts to realize aligned natural inflation mainly focused on type IIB orientifold compactifications with fluxes, where moduli stabilization has been extensively studied. In these constructions, the  \(G_3\) flux induces a superpotential~\cite{Gukov:1999ya}
\begin{equation}
W_{\text{flux}}=\intop_{Y}G_3\wedge \Omega\,,
\end{equation}
where \(Y\) labels the complex three dimensional space and \(\Omega\) is the holomorphic 3-form of \(Y\). This superpotential might be sufficient to stabilize all moduli~\cite{Giddings:2001yu} except for the K\"ahler moduli which do not enter \(W_{\text{flux}}\). Assuming that this stabilization takes place at a high scale, the heavy moduli can be integrated out, resulting in an effectively constant superpotential \(W_{\text{flux}}=W_0\). Non-perturbative effects then generate a superpotential for the K\"ahler moduli \(T_{\beta}\)~\cite{Kachru:2003aw}
\begin{equation}\label{eq:type2}
W=W_{\text{flux}}+W_{\text{NP}}=W_0+\sum_iA_ie^{-2\pi n_i^{\beta}T_{\beta}}\,.
\end{equation}
Here \(A_i\) is a model-dependent prefactor and the matrix \(n_i^{\beta}\) depends on the microscopic origin of the non-perturbative effect. The latter may stem from gaugino condensates on D5 or D7 branes, Euclidean D1 or D3 brane instantons~\cite{Grana:2005jc,Blumenhagen:2006ci}. The axionic components of the K\"ahler moduli obtain periodic potentials which could trigger natural inflation (see~\cite{Grimm:2007hs} for a comprehensive discussion).

Indeed, models with several K\"ahler moduli and multiple gaugino condensates have been constructed. In combination with the constant flux-induced superpotential and an uplifting sector they can lead to the desired potential for aligned natural inflation~\cite{Long:2014dta,Ben-Dayan:2014zsa,Ben-Dayan:2014lca}. These models are generalizations of the supergravity model in section~\ref{sec:highm32}. Similar to their supergravity counterpart, they require a very high supersymmetry breaking scale~\cite{Long:2014dta}. The reason is that the inflationary scale in these schemes is set by $W_0$ (see section~\ref{sec:highm32}).\footnote{
Another reason for the large gravitino mass is that the potential barrier which protects the saxionic part of the K\"ahler moduli during inflation is intimately related to $m_{3/2}$. Stable minima during inflation enforce $m_{3/2}\gg H \sim 10^{-4}$.}

On the contrary, high scale supersymmetry breaking can be avoided if the inflationary scale is unrelated to $m_{3/2}$. As outlined in section~\ref{sec:lowscale} this can be achieved by the inclusion of one or several matter fields which act as stabilizers. It is well known that the prefactor of Euclidean D brane instanton contributions can depend on other moduli as well as chiral matter~\cite{Grimm:2007xm,Blumenhagen:2006xt,Ibanez:2006da,Florea:2006si}. Instead of neglecting this dependence, it can lead to a natural origin of stabilizer fields in string theory. 

The non-perturbative superpotential contribution including matter fields \(\phi_i\) can schematically be written as~\cite{Blumenhagen:2009qh}
\begin{equation}
\label{eq:instanton}
W_\text{NP} = \prod_i \phi_i e^{-S_\text{inst}(T_{\beta})}\,,
\end{equation}
where $S_\text{inst}$ denotes the instanton action. The inclusion of matter fields in these constructions is highly non-trivial. For this reason we shall for the moment turn to the simpler toroidal
compactifications, where effects are better understood and explicit calculations exist. To deal with concrete numbers we will concentrate on world-sheet instantons as origin of the non-perturbative superpotential. As an example, we consider the case of magnetized D9 branes for a factorizable toroidal compactification \(\mathbbm{T}^6=\bigotimes_{\beta=1}^3\mathbbm{T}^2_{\beta}\). Couplings between three chiral matter fields can be induced by world-sheet instantons, which in type IIB compactifications depend on the complex structure moduli \(U_{\beta}\). They are given as~\cite{Blumenhagen:2006ci}
\begin{equation}
W_{\text{NP}} = \phi_1 \phi_2\phi_3 A e^{-\pi \kappa^{\beta}U_{\beta}}\,,\qquad 
\kappa^{\beta}=
\left(\frac{i^{\beta}}{I_{ab}^{\beta}}+\frac{j^{\beta}}{I_{bc}^{\beta}}+\frac{k^{\beta}}{I_{ca}^{\beta}}\right)|I_{ab}^{\beta}I_{bc}^{\beta}I_{ca}^{\beta}|\,.
\end{equation}
Here we have already approximated the \(\vartheta\)-function by its leading order\footnote{Recently it was outlined that \(\vartheta\)-functions may lead to interesting signatures if they serve as potentials for inflation~\cite{Higaki:2015kta}.}. The parameters \(\kappa^{\beta}\) depend on the intersection numbers \(I_{ab}^{\beta}\), \(I_{bc}^{\beta}\), \(I_{ca}^{\beta}\) and intersection points \(i^{\beta},j^{\beta},k^{\beta}\) and are in general fractional numbers. The intersection numbers are given by discrete flux numbers \(I_{ab}^{\beta}=p_a^{\beta}q_b^{\beta}-p_b^{\beta}q_a^{\beta}\). These numbers can differ for each subtorus contribution resulting in a different exponential dependence on the three moduli \(U_{\beta}\). The prefactor \(A\) is independent of the moduli \(U_{\beta}\).

In this setup, one or several matter fields could play the role of the stabilizer, while the inflaton is identified with a linear combination of the axionic parts of the $U_\beta$. A near alignment of axion decay constants as required in the KNP mechanism can result by the choice of suitable intersection numbers. While we considered toroidal compactifications as a concrete example we expect similar contributions in compactifications on more sophisticated backgrounds. Analogous ideas may also be employed in the heterotic string, where world-sheet instantons can induce similar non-perturbative terms including matter fields (see e.g.~\cite{Dundee:2010sb}).

In order to avoid high scale supersymmetry breaking we further have to avoid large \(W_0\). This can be achieved if the flux contribution to \(W_0\) is absent or tuned to be small~\cite{Kachru:2003aw}. Alternatively, a large flux contribution can be canceled by moduli contributions to $W_0$~\cite{Kallosh:2004yh,Krefl:2006vu}. 

\section{Aligned Natural Inflation in a String-inspired Model}\label{sec:model}
In the following, we will implement the KNP alignment mechanism in a concrete string-inspired model with instantons. 

\subsection{The Model}
The scheme, we wish to advertise, bears resemblance with the natural inflation model with stabilizer discussed in section~\ref{sec:lowscale}. However, the inflaton will be identified with a linear combination of two axions allowing for an effective trans-Planckian axion decay constant. We consider the superpotential and K\"ahler potential
\begin{subequations}\label{eq:stringymodel}
\begin{align}
W &= \sum_{i=1}^2 m_i^2 X_i\, (e^{-a_i\rho_1 - b_i \rho_2} - \lambda_i)\,,\\
K&= -\sum_{i=1}^2 \log (\bar{\rho}_i + \rho_i) + k(|X_i|^2)\,,
\end{align}
\end{subequations}
which contains the matter fields $X_{1,2}$ as well as the moduli $\rho_{1,2}$.
The non-perturbative terms in the superpotential above can result from world-sheet instantons or D brane instantons as outlined in section~\ref{sec:instantons}.
The $m_{1,2}$ and $\lambda_{1,2}$ are in general functions of further chiral matter fields and moduli which we assume to be stabilized with non-trivial vacuum expectation values. Taking the stabilization scale to be sufficiently high, we can treat them as effective constants.\footnote{We assume that $m_{1,2}$ and $\lambda_{1,2}$ are real as phases can be absorbed by appropriate field redefinitions.} The K\"ahler potentials of $X_{1,2}$ may carry moduli dependences which, however, do not play a role for the following discussion.

One easily verifies that this model has a supersymmetric ground state at $X_{1,2}=0$ and
\begin{equation}
\rho_1= \frac{b_2 \log\lambda_1-b_1\log\lambda_2}{a_2 b_1-a_1 b_2} \equiv \rho_{1,0}\,,\qquad 
\rho_2= \frac{a_2 \log\lambda_1-a_1\log\lambda_2}{b_2 a_1-b_1 a_2} \equiv \rho_{2,0}\,.
\end{equation}
Similar as in~\cite{Wieck:2014xxa}, moduli are stabilized by the F-terms of matter fields and hence their mass is unrelated to the gravitino mass. In order to simplify the analytic discussion, let us assume that $\lambda_2 m_2^2 \gg \lambda_1 m_1^2$. In this case one linear combination of moduli $a_2 \rho_1 + b_2 \rho_2$ is heavy, while the orthogonal linear combination remains light. We redefine
\begin{equation}
\rho_1= \rho_{1,0} + b_2 \rho + a_2 \tilde{\rho}\,, \qquad \rho_2= \rho_{2,0} -a_2 \rho + b_2 \tilde{\rho}\,,
\end{equation}
where $\tilde{\rho}$ and $\rho$ correspond to the heavy and light linear combination of moduli (up to a normalization factor and a shift by the vacuum expectation values). 
Now we set the heavy fields $X_{1,2}$ and $\tilde{\rho}$ to their minima and decompose $\rho=\chi + i \varphi$ into saxion $\chi$ and axion $\varphi$.
The potential becomes 
\begin{equation}\label{eq:axsax}
V = \frac{\lambda_1^2 m_1^4 \,e^{-\delta \chi} \left[ \cosh(\delta\chi) - \cos (\delta \varphi) \right]}{2 (\rho_{1,0} + b_2 \chi) (\rho_{2,0} - a_2 \chi)}\,,
\end{equation}
where we introduced
\begin{equation}
\delta = a_1 b_2 - a_2 b_1\,.
\end{equation}
This potential resembles~\eqref{eq:modpot} with two major differences: first, a trans-Planckian axion decay constant can arise in the case of near alignment $a_1b_2 \sim a_2 b_1$, where $\delta$ is suppressed. Secondly, the potential now has two poles at $\chi= -\rho_{1,0}/b_2$ and $\chi=\rho_{2,0}/a_2$ between which the saxion is trapped. These poles result from $\rho$ being a linear combination of two moduli. They correspond to the poles at $\rho_{1,2}=0$ in terms of the original fields. We note that the effective description~\eqref{eq:axsax} breaks down above the stabilization scale of the heavy linear combination $\tilde{\rho}$. But as long as $m_{\tilde{\rho}} > H$ is fulfilled,~\eqref{eq:axsax} is a valid approximation. 

\begin{figure}[t]
\begin{center}
\includegraphics[height=10.5cm]{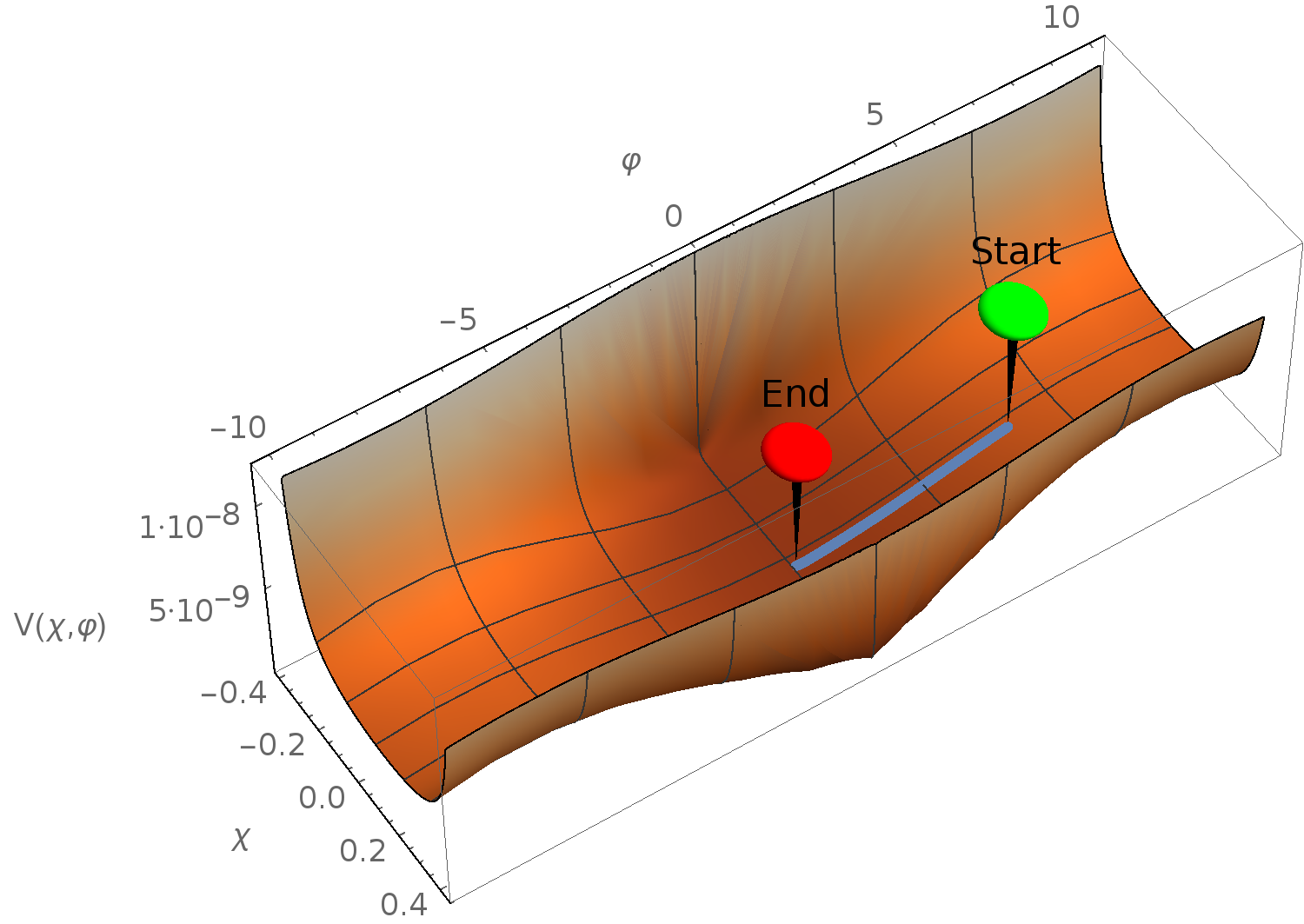}\qquad
\end{center}
\caption{Potential~\eqref{eq:axsax} in axion and saxion direction. The inflationary trajectory (last 60 e-folds) is indicated by the gray line. Observe that we have stretched the scale of $\chi$ compared to $\varphi$. The valley is in fact much narrower than it appears in the figure.}
\label{fig:inflation3d}
\end{figure}

\begin{table}[ht]
\begin{center}
\begin{tabular}{|c|c|c|c|c|c|c|c||c|c|c|c|}
\hline
$a_1$ & $b_1$ & $a_2$ & $b_2$ &$m_1$ & $m_2$ & $\lambda_1$ & $\lambda_2$ & $\rho_{1,0}$ &$\rho_{2,0}$ & $\delta$ & $f_\text{eff}$ \\
\hline\hwl
$\frac{7\pi}{6}$ & $\frac{8\pi}{7}$ & $\frac{6\pi}{7}$ & $\frac{7\pi}{8}$ &$0.45$ & $0.55$ & $2.6\cdot 10^{-4}$ & $2.0\cdot 10^{-3}$ & $1.1$ & $1.2$ & $0.41$& $5.9$\\[0.2mm]
\hline
\end{tabular}
\end{center}
\caption{Exemplary parameter choice which leads to a successful realization of aligned natural inflation.}
\label{tab:benchmark}
\end{table}

\begin{figure}[t]
\begin{center}
\includegraphics[height=6.5cm]{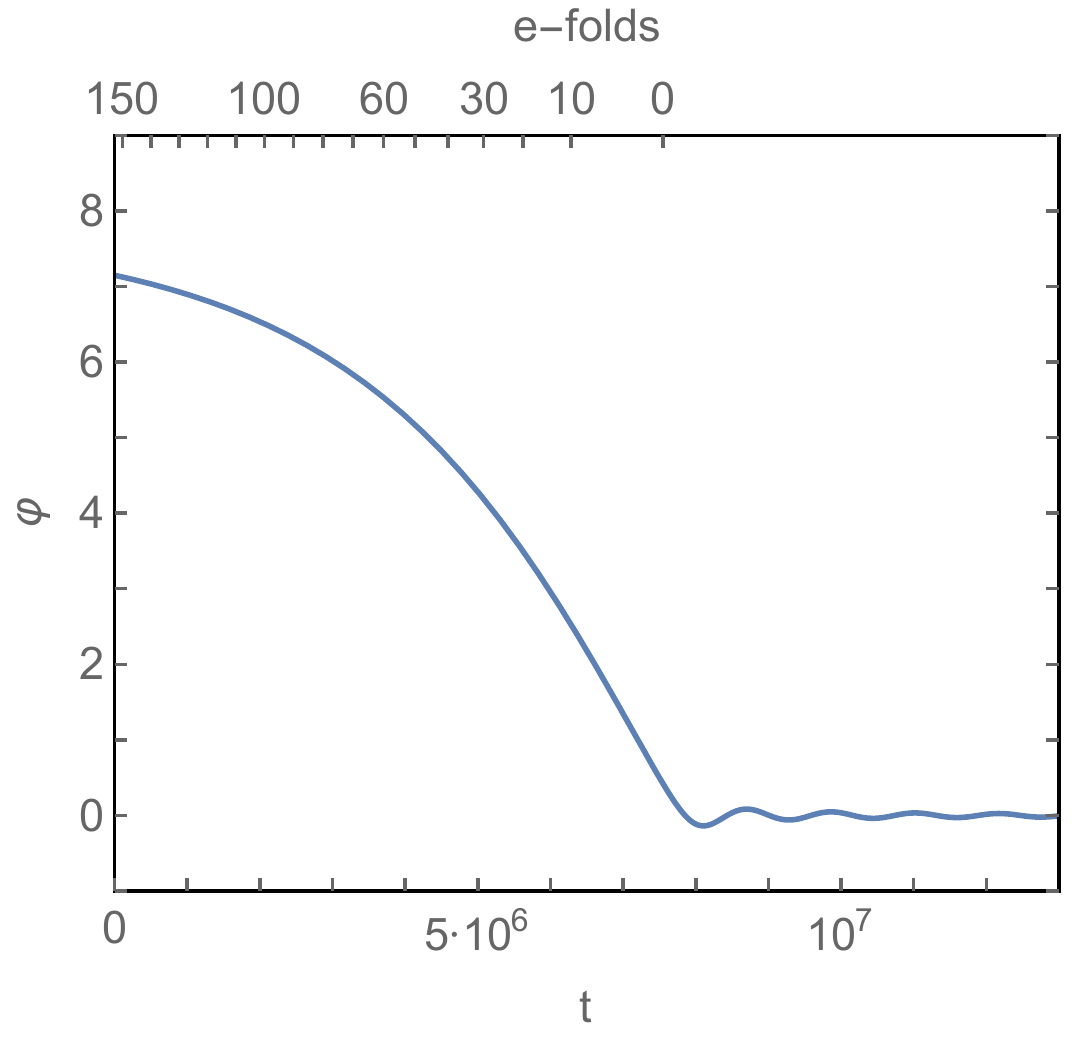}\qquad
\includegraphics[height=6.5cm]{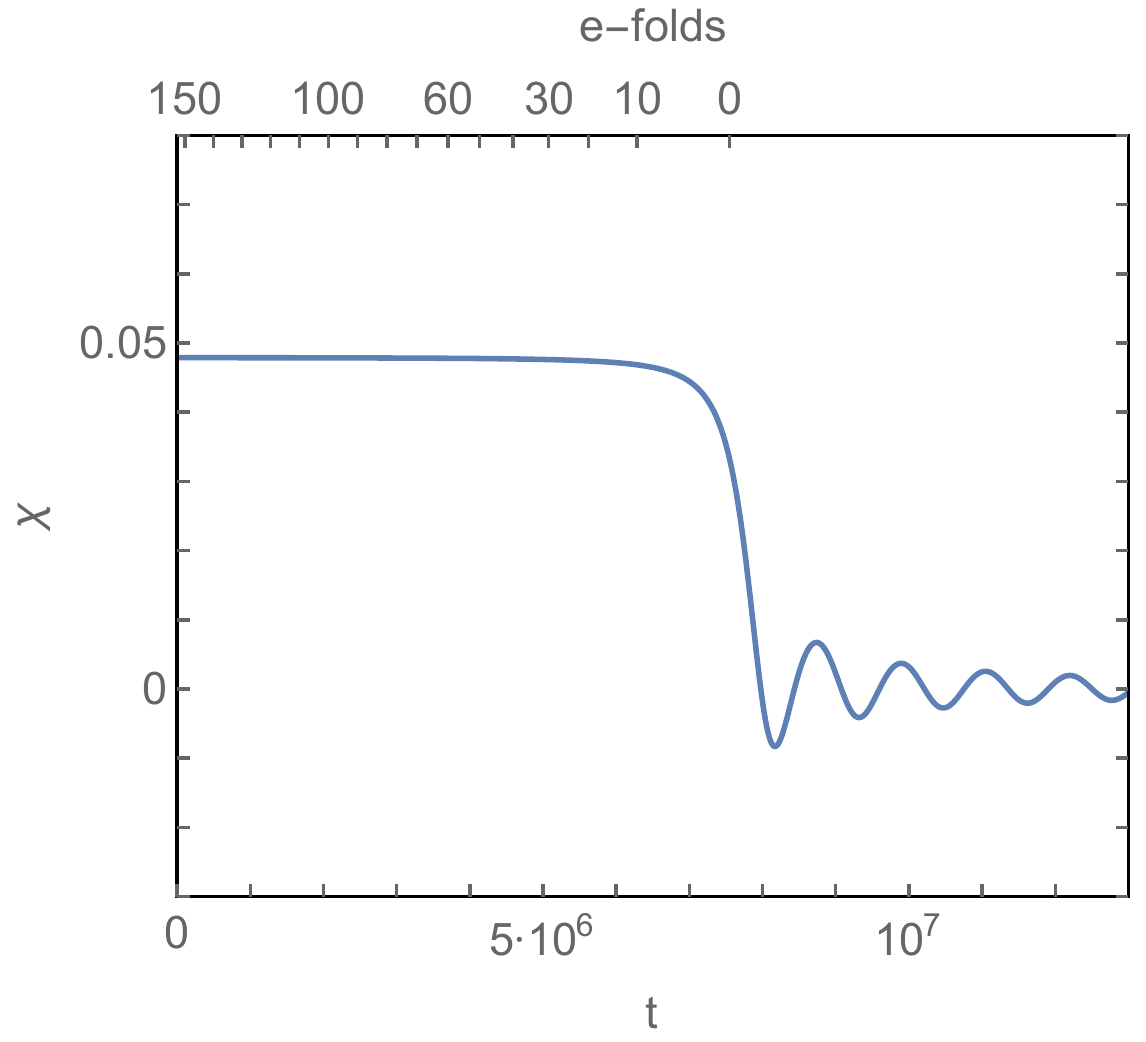}
\end{center}
\caption{Axion (left) and saxion (right) displacement as a function of time. Also shown is the corresponding number of e-folds. During inflation, the saxion is frozen at a non-vanishing field value by a large Hubble mass term.}
\label{fig:fieldtrajectories}
\end{figure}

In figure~\ref{fig:inflation3d}, we depict the potential along the light axion and saxion directions for the parameter choice of table~\ref{tab:benchmark}. At the origin of field space, both fields are mass-degenerate due to unbroken supersymmetry. However, slightly away from the minimum, the saxion potential becomes very steep. The inflaton is identified with the axion which resides in a periodic potential. Along the inflationary trajectory, which is also indicated in figure~\ref{fig:inflation3d}, the potential reduces to
\begin{equation}
V \simeq \Lambda^4\; \left[1-\cos(\delta \varphi)\right]\,,\qquad \Lambda^4=\frac{\lambda_1^2m_1^4 \,e^{-\delta \chi} }{2 (\rho_{1,0} + b_2 \chi) (\rho_{2,0} - a_2 \chi)}\,.
\end{equation}
During inflation $\chi$ remains fixed at the minimum of the function $\Lambda^4$ and receives a large Hubble mass term. Hence, we can treat $\chi$ as a constant. Only at the end of inflation the field returns to its minimum at $\chi=0$. This can be seen in figure~\ref{fig:fieldtrajectories}, where we depict $\chi$ and $\varphi$ as a function of time.

We have to take into account that the axion is not yet canonically normalized. The axion kinetic term reads $\mathcal{L}_\text{kin}= G_{\varphi\varphi} \,\partial_\mu \varphi \partial^\mu \varphi$. The prefactor $G_{\varphi\varphi}$ is related to the K\"ahler metrics of the moduli
\begin{equation}
G_{\varphi\varphi}  =b_2^2\, K_{\rho_1\bar{\rho}_1} -a_2^2\, K_{\rho_2\bar{\rho}_2} = \frac{b_2^2}{4(\rho_{1,0} + b_2 \chi)^2}+\frac{a_2^2}{4(\rho_{2,0} - a_2 \chi)^2}\,.
\end{equation}
The effective axion decay constant including the normalization of the kinetic term reads
\begin{equation}
f_\text{eff} = \frac{\sqrt{2 G_{\varphi\varphi} }}{\delta}\simeq \frac{1}{\delta}\:\sqrt{\frac{b_2^2}{2\rho_{1,0}^2}+\frac{a_2^2}{2\rho_{2,0}^2}}\,,
\end{equation}
where the second equality holds if the saxion shift is negligible.
In the example of table~\ref{tab:benchmark} this approximation is valid and we obtain $f_\text{eff}=5.9$. The corresponding CMB observables for 60 e-folds of inflation are given as $n_s=0.959$ and $r=0.050$.

\subsection{Comparison with Observation}

\begin{figure}[t]
\begin{center}
\includegraphics[height=7.5cm]{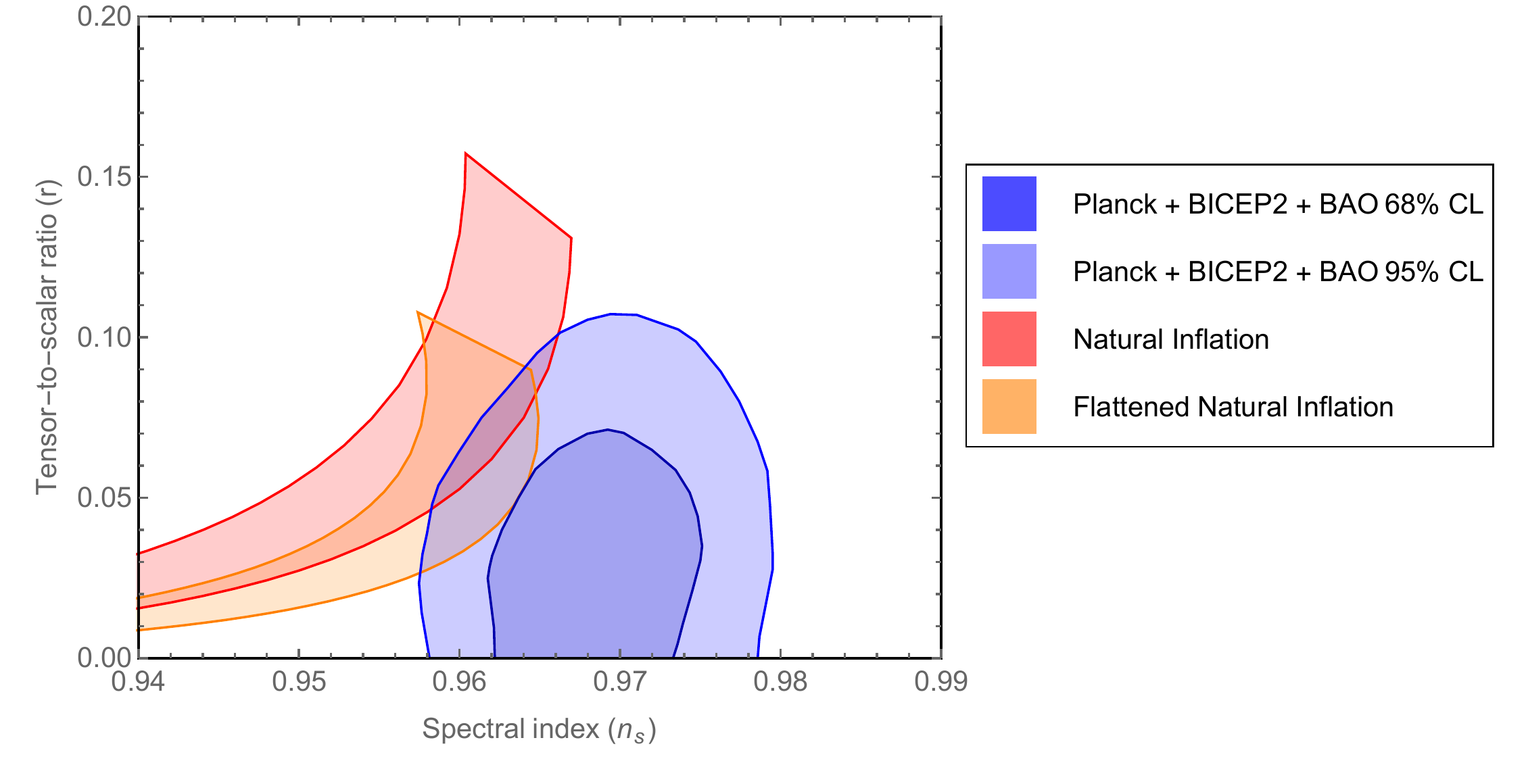}
\end{center}
\caption{Combined Planck, BICEP2/Keck Array, BAO constraints on the spectral index and the tensor-to-scalar ratio~\cite{Ade:2015lrj}. Predictions of natural inflation with and without including flattening effects (see text) are also shown. The upper and lower ends of the bands refer to 50 and 60 e-folds respectively.}
\label{fig:flatten}
\end{figure}

The model we presented realizes natural inflation. The corresponding predictions for the CMB observables $n_s$ and $r$ are depicted in figure~\ref{fig:flatten}. Also shown are the most recent Planck constraints cross-correlated with BICEP2/Keck Array and baryon acoustic oscillation (BAO) data~\cite{Ade:2015lrj}. If we take the 95\% confidence level contour as a reference, an (effective) axion decay constant
\begin{equation}
f_\text{eff}= 5.8- 10.9
\end{equation}
 is required to match observations.

Let us now briefly discuss corrections to the inflaton potential which could affect the CMB observables. String compactifications generically result in a plethora of scalar fields including moduli as well as chiral matter fields which could potentially affect the inflationary potential (see~\cite{Dong:2010in}). To be specific let us add to~\eqref{eq:stringymodel} a heavy modulus $\sigma$ with a K\"ahler potential $K=-\log(\bar{\sigma}+\sigma)$ without direct couplings to the inflaton sector. Assuming the modulus to be stabilized supersymmetrically with a mass $M>H$, the heavy modulus induces a correction~\cite{Buchmuller:2014vda}
\begin{equation}\label{eq:flattening}
V= V_0 \left( 1 - \frac{V_0}{M^2}\right)
\end{equation}
to the inflaton potential at leading order in $V_0/M^2$. Here $V_0$ denotes the inflaton potential without the heavy modulus. If several heavy moduli or matter fields induce corrections, their contributions can be added leading to a smaller effective $M$ in~\eqref{eq:flattening}. The above described effect leads to a flattening of the inflaton potential and hence reduces the amplitude in gravitational waves. In figure~\ref{fig:flatten} we also depict the ($n_s$,$r$)-band for natural inflation including the flattening effect. We chose $M^2=V_0^*/5$, where $V_0^*$ denotes the value of $V_0$ at horizon crossing of the scales relevant for the CMB observables. As can be seen, the flattening effect improves the agreement with the Planck data.

So far we have concentrated on the correction from heavy scalars without direct couplings to the inflaton. However, we also expect corrections from further non-perturbative contributions involving the moduli $\rho_{1,2}$ of the inflation sector. These result from expanding the $\vartheta$-function to higher orders or from additional string instanton couplings in the superpotential. Such terms could induce modulations to the inflaton potential very similar as in axion monodromy~\cite{Silverstein:2008sg} which affect the CMB observables and lead to distinctive features in the CMB~\cite{Flauger:2009ab}. We leave a more detailed analysis of this possibility for future work.

\section{Conclusions}

In the present paper we have shown that low scale supersymmetry
is compatible with high scale inflation in the framework of
superstring theory. Apart from the inflaton multiplet, the
mechanism requires the presence of one additional
modulus. This additional field solves two problems simultaneously:
first it provides an effective trans-Planckian axion decay
constant via the alignment mechanism of KNP and secondly it
leads to an inflationary potential with a ``trapped saxion''
(see figure~\ref{fig:inflation3d}). In section~\ref{sec:model} we discussed a specific string-inspired
model of aligned natural inflation. This model possesses a supersymmetric ground state where moduli are stabilized through F-terms of matter fields. Non-perturbative contributions
to the potential originate from world-sheet instantons on
wrapped brane configurations. A choice of suitable intersection
numbers allows a satisfactory alignment and an effective
trans-Planckian decay constant of the lightest axion. This leads
to natural inflation with a single effective axion. Predictions
for the spectral index ($n_s$) versus tensor-to-scalar ratio
($r$) are shown in figure~\ref{fig:flatten}. Experiments in the coming years should
be able to confirm or falsify the scheme.

\subsection*{Acknowledgments}

We would like to thank Stefan F\"orste and Hans Jockers for helpful discussions. This work was supported by the SFB-Transregio TR33 "The Dark Universe" 
(Deutsche Forschungsgemeinschaft).

%%%%%%%%%%%%%%%%%%%%%%%%%%%%%%%%%%%%%%%%%%%%%%%%%%%%%%%%%%%%%%%%%%%%%%%
\bibliography{natural}

\providecommand{\bysame}{\leavevmode\hbox to3em{\hrulefill}\thinspace}
\begin{thebibliography}{10}

\bibitem{Lyth:1996im}
D.~H. Lyth, Phys.Rev.Lett. \textbf{78} (1997), 1861--1863,  [hep-ph/9606387].
%%CITATION = HEP-PH/9606387;%%

\bibitem{Ade:2014xna}
BICEP2 Collaboration, P.~Ade et~al., Phys.Rev.Lett. \textbf{112} (2014),
  no.~24, 241101,  [1403.3985].
%%CITATION = ARXIV:1403.3985;%%

\bibitem{Ade:2015tva}
BICEP2 Collaboration, Planck Collaboration, P.~Ade et~al., Phys.Rev.Lett.
  (2015),  [1502.00612].
%%CITATION = ARXIV:1502.00612;%%

\bibitem{Freese:1990rb}
K.~Freese, J.~A. Frieman, and A.~V. Olinto, Phys.Rev.Lett. \textbf{65} (1990),
  3233--3236.
%%CITATION = PRLTA,65,3233;%%

\bibitem{Kim:2004rp}
J.~E. Kim, H.~P. Nilles, and M.~Peloso, JCAP \textbf{0501} (2005), 005,
  [hep-ph/0409138].
%%CITATION = HEP-PH/0409138;%%

\bibitem{Silverstein:2008sg}
E.~Silverstein and A.~Westphal, Phys.Rev. \textbf{D78} (2008), 106003,
  [0803.3085].
%%CITATION = ARXIV:0803.3085;%%

\bibitem{Kallosh:2007cc}
R.~Kallosh, N.~Sivanandam, and M.~Soroush, Phys.Rev. \textbf{D77} (2008),
  043501,  [0710.3429].
%%CITATION = ARXIV:0710.3429;%%

\bibitem{Czerny:2014xja}
M.~Czerny, T.~Higaki, and F.~Takahashi, JHEP \textbf{1405} (2014), 144,
  [1403.0410].
%%CITATION = ARXIV:1403.0410;%%

\bibitem{Long:2014dta}
C.~Long, L.~McAllister, and P.~McGuirk, Phys.Rev. \textbf{D90} (2014), no.~2,
  023501,  [1404.7852].
%%CITATION = ARXIV:1404.7852;%%

\bibitem{Ben-Dayan:2014zsa}
I.~Ben-Dayan, F.~G. Pedro, and A.~Westphal, Phys.Rev.Lett. \textbf{113} (2014),
  no.~26, 261301,  [1404.7773].
%%CITATION = ARXIV:1404.7773;%%

\bibitem{Kallosh:2014vja}
R.~Kallosh, A.~Linde, and B.~Vercnocke, Phys.Rev. \textbf{D90} (2014), no.~4,
  041303,  [1404.6244].
%%CITATION = ARXIV:1404.6244;%%

\bibitem{Gao:2014uha}
X.~Gao, T.~Li, and P.~Shukla, JCAP \textbf{1410} (2014), no.~10, 048,
  [1406.0341].
%%CITATION = ARXIV:1406.0341;%%

\bibitem{Li:2014lpa}
T.~Li, Z.~Li, and D.~V. Nanopoulos, JHEP \textbf{1411} (2014), 012,
  [1407.1819].
%%CITATION = ARXIV:1407.1819;%%

\bibitem{Ben-Dayan:2014lca}
I.~Ben-Dayan, F.~G. Pedro, and A.~Westphal,  (2014),  1407.2562.
%%CITATION = ARXIV:1407.2562;%%

\bibitem{Abe:2014pwa}
H.~Abe, T.~Kobayashi, and H.~Otsuka,  (2014),  1409.8436.
%%CITATION = ARXIV:1409.8436;%%

\bibitem{Ali:2014mra}
T.~Ali, S.~S. Haque, and V.~Jejjala,  (2014),  1410.4660.
%%CITATION = ARXIV:1410.4660;%%

\bibitem{Abe:2014xja}
H.~Abe, T.~Kobayashi, and H.~Otsuka,  (2014),  1411.4768.
%%CITATION = ARXIV:1411.4768;%%

\bibitem{Bachlechner:2014gfa}
T.~C. Bachlechner, C.~Long, and L.~McAllister,  (2014),  1412.1093.
%%CITATION = ARXIV:1412.1093;%%

\bibitem{Li:2014unh}
T.~Li, Z.~Li, and D.~V. Nanopoulos,  (2014),  1412.5093.
%%CITATION = ARXIV:1412.5093;%%

\bibitem{Garcia-Etxebarria:2014wla}
I.~García-Etxebarria, T.~W. Grimm, and I.~Valenzuela,  (2014),  1412.5537.
%%CITATION = ARXIV:1412.5537;%%

\bibitem{Ade:2015lrj}
Planck Collaboration, P.~Ade et~al.,  (2015),  1502.02114.
%%CITATION = ARXIV:1502.02114;%%

\bibitem{Buchmuller:2015oma}
W.~Buchmuller, E.~Dudas, L.~Heurtier, A.~Westphal, C.~Wieck, and M.~W. Winkler,
   (2015),  1501.05812.
%%CITATION = ARXIV:1501.05812;%%

\bibitem{Kawasaki:2000yn}
M.~Kawasaki, M.~Yamaguchi, and T.~Yanagida, Phys.Rev.Lett. \textbf{85} (2000),
  3572--3575,  [hep-ph/0004243].
%%CITATION = HEP-PH/0004243;%%

\bibitem{Kallosh:2010ug}
R.~Kallosh and A.~Linde, JCAP \textbf{1011} (2010), 011,  [1008.3375].
%%CITATION = ARXIV:1008.3375;%%

\bibitem{Kallosh:2010xz}
R.~Kallosh, A.~Linde, and T.~Rube, Phys.Rev. \textbf{D83} (2011), 043507,
  [1011.5945].
%%CITATION = ARXIV:1011.5945;%%

\bibitem{Polonyi:1977}
J.~Polonyi, Hungary Central Inst. Res. preprint \textbf{KFKI-77-93} (1977).

\bibitem{Coughlan:1983ci}
G.~Coughlan, W.~Fischler, E.~W. Kolb, S.~Raby, and G.~G. Ross, Phys.Lett.
  \textbf{B131} (1983), 59.
%%CITATION = PHLTA,B131,59;%%

\bibitem{Dine:1983ys}
M.~Dine, W.~Fischler, and D.~Nemeschansky, Phys.Lett. \textbf{B136} (1984),
  169.
%%CITATION = PHLTA,B136,169;%%

\bibitem{Coughlan:1984yk}
G.~Coughlan, R.~Holman, P.~Ramond, and G.~G. Ross, Phys.Lett. \textbf{B140}
  (1984), 44.
%%CITATION = PHLTA,B140,44;%%

\bibitem{Wen:1985jz}
X.~Wen and E.~Witten, Phys.Lett. \textbf{B166} (1986), 397.
%%CITATION = PHLTA,B166,397;%%

\bibitem{Dine:1986vd}
M.~Dine and N.~Seiberg, Phys.Rev.Lett. \textbf{57} (1986), 2625.
%%CITATION = PRLTA,57,2625;%%

\bibitem{Marchesano:2014mla}
F.~Marchesano, G.~Shiu, and A.~M. Uranga, JHEP \textbf{1409} (2014), 184,
  [1404.3040].
%%CITATION = ARXIV:1404.3040;%%

\bibitem{Hebecker:2014eua}
A.~Hebecker, S.~C. Kraus, and L.~T. Witkowski, Phys.Lett. \textbf{B737} (2014),
  16--22,  [1404.3711].
%%CITATION = ARXIV:1404.3711;%%

\bibitem{Blumenhagen:2014nba}
R.~Blumenhagen, D.~Herschmann, and E.~Plauschinn, JHEP \textbf{1501} (2015),
  007,  [1409.7075].
%%CITATION = ARXIV:1409.7075;%%

\bibitem{Banks:2003sx}
T.~Banks, M.~Dine, P.~J. Fox, and E.~Gorbatov, JCAP \textbf{0306} (2003), 001,
  [hep-th/0303252].
%%CITATION = HEP-TH/0303252;%%

\bibitem{Svrcek:2006yi}
P.~Svrcek and E.~Witten, JHEP \textbf{0606} (2006), 051,  [hep-th/0605206].
%%CITATION = HEP-TH/0605206;%%

\bibitem{Blumenhagen:2014gta}
R.~Blumenhagen and E.~Plauschinn, Phys.Lett. \textbf{B736} (2014), 482--487,
  [1404.3542].
%%CITATION = ARXIV:1404.3542;%%

\bibitem{Grimm:2014vva}
T.~W. Grimm, Phys.Lett. \textbf{B739} (2014), 201--208,  [1404.4268].
%%CITATION = ARXIV:1404.4268;%%

\bibitem{Witten:1985xb}
E.~Witten, Phys.Lett. \textbf{B155} (1985), 151.
%%CITATION = PHLTA,B155,151;%%

\bibitem{Buchmuller:2004xr}
W.~Buchmuller, K.~Hamaguchi, O.~Lebedev, and M.~Ratz, Nucl.Phys. \textbf{B699}
  (2004), 292--308,  [hep-th/0404168].
%%CITATION = HEP-TH/0404168;%%

\bibitem{Kallosh:2004yh}
R.~Kallosh and A.~D. Linde, JHEP \textbf{0412} (2004), 004,  [hep-th/0411011].
%%CITATION = HEP-TH/0411011;%%

\bibitem{Tye:2014tja}
S.~H.~H. Tye and S.~S.~C. Wong,  (2014),  1404.6988.
%%CITATION = ARXIV:1404.6988;%%

\bibitem{Kappl:2014lra}
R.~Kappl, S.~Krippendorf, and H.~P. Nilles, Phys.Lett. \textbf{B737} (2014),
  124--128,  [1404.7127].
%%CITATION = ARXIV:1404.7127;%%

\bibitem{Choi:2014rja}
K.~Choi, H.~Kim, and S.~Yun, Phys.Rev. \textbf{D90} (2014), no.~2, 023545,
  [1404.6209].
%%CITATION = ARXIV:1404.6209;%%

\bibitem{Gukov:1999ya}
S.~Gukov, C.~Vafa, and E.~Witten, Nucl.Phys. \textbf{B584} (2000), 69--108,
  [hep-th/9906070].
%%CITATION = HEP-TH/9906070;%%

\bibitem{Giddings:2001yu}
S.~B. Giddings, S.~Kachru, and J.~Polchinski, Phys.Rev. \textbf{D66} (2002),
  106006,  [hep-th/0105097].
%%CITATION = HEP-TH/0105097;%%

\bibitem{Kachru:2003aw}
S.~Kachru, R.~Kallosh, A.~D. Linde, and S.~P. Trivedi, Phys.Rev. \textbf{D68}
  (2003), 046005,  [hep-th/0301240].
%%CITATION = HEP-TH/0301240;%%

\bibitem{Grana:2005jc}
M.~Grana, Phys.Rept. \textbf{423} (2006), 91--158,  [hep-th/0509003].
%%CITATION = HEP-TH/0509003;%%

\bibitem{Blumenhagen:2006ci}
R.~Blumenhagen, B.~Kors, D.~Lust, and S.~Stieberger, Phys.Rept. \textbf{445}
  (2007), 1--193,  [hep-th/0610327].
%%CITATION = HEP-TH/0610327;%%

\bibitem{Grimm:2007hs}
T.~W. Grimm, Phys.Rev. \textbf{D77} (2008), 126007,  [0710.3883].
%%CITATION = ARXIV:0710.3883;%%

\bibitem{Grimm:2007xm}
T.~W. Grimm, JHEP \textbf{0710} (2007), 004,  [0705.3253].
%%CITATION = ARXIV:0705.3253;%%

\bibitem{Blumenhagen:2006xt}
R.~Blumenhagen, M.~Cvetic, and T.~Weigand, Nucl.Phys. \textbf{B771} (2007),
  113--142,  [hep-th/0609191].
%%CITATION = HEP-TH/0609191;%%

\bibitem{Ibanez:2006da}
L.~Ibanez and A.~Uranga, JHEP \textbf{0703} (2007), 052,  [hep-th/0609213].
%%CITATION = HEP-TH/0609213;%%

\bibitem{Florea:2006si}
B.~Florea, S.~Kachru, J.~McGreevy, and N.~Saulina, JHEP \textbf{0705} (2007),
  024,  [hep-th/0610003].
%%CITATION = HEP-TH/0610003;%%

\bibitem{Blumenhagen:2009qh}
R.~Blumenhagen, M.~Cvetic, S.~Kachru, and T.~Weigand, Ann.Rev.Nucl.Part.Sci.
  \textbf{59} (2009), 269--296,  [0902.3251].
%%CITATION = ARXIV:0902.3251;%%

\bibitem{Higaki:2015kta}
T.~Higaki and F.~Takahashi,  (2015),  1501.02354.
%%CITATION = ARXIV:1501.02354;%%

\bibitem{Dundee:2010sb}
B.~Dundee, S.~Raby, and A.~Westphal, Phys.Rev. \textbf{D82} (2010), 126002,
  [1002.1081].
%%CITATION = ARXIV:1002.1081;%%

\bibitem{Krefl:2006vu}
D.~Krefl and D.~Lust, JHEP \textbf{0606} (2006), 023,  [hep-th/0603166].
%%CITATION = HEP-TH/0603166;%%

\bibitem{Wieck:2014xxa}
C.~Wieck and M.~W. Winkler, Phys.Rev. \textbf{D90} (2014), no.~10, 103507,
  [1408.2826].
%%CITATION = ARXIV:1408.2826;%%

\bibitem{Dong:2010in}
X.~Dong, B.~Horn, E.~Silverstein, and A.~Westphal, Phys.Rev. \textbf{D84}
  (2011), 026011,  [1011.4521].
%%CITATION = ARXIV:1011.4521;%%

\bibitem{Buchmuller:2014vda}
W.~Buchmuller, C.~Wieck, and M.~W. Winkler, Phys.Lett. \textbf{B736} (2014),
  237--240,  [1404.2275].
%%CITATION = ARXIV:1404.2275;%%

\bibitem{Flauger:2009ab}
R.~Flauger, L.~McAllister, E.~Pajer, A.~Westphal, and G.~Xu, JCAP \textbf{1006}
  (2010), 009,  [0907.2916].
%%CITATION = ARXIV:0907.2916;%%

\end{thebibliography}
\bibliographystyle{ArXiv}

\end{document}